\documentstyle[12pt]{article}
\def\be {\begin{equation}}
\def\ee {\end{equation}}
\def\ba {\begin{eqnarray}}
\def\ea {\end{eqnarray}}

\def\bi {\begin{itemize}}
\def\ei {\end{itemize}}
\begin{document}
\def\bea{\begin{eqnarray}}
\def\eea{\end{eqnarray}}
\title{\bf  {Interacting  Dark Energy in  Ho\v{r}ava-Lifshitz Cosmology}}
\author{M.R. Setare  \footnote{E-mail: rezakord@ipm.ir}
  \\ {Department of Science,  Payame Noor University. Bijar, Iran}}
\date{\small{}}

\maketitle
\begin{abstract}
In the usual Ho\v{r}ava-Lifshitz cosmological models, the scalar
field is responsible for dark matter. Using an additional scalar
field, Saridakis \cite{sari} has formulated Ho\v{r}ava-Lifshitz
cosmology with an effective dark energy sector. In the paper
\cite{sari} the scalar fields do not interact with each other,
here we extend this work to the interacting case, where matter
scalar field $\phi$ interact with dark energy scalar field
$\sigma$. We will show that in contrast with \cite{sari}, where
$\sigma$-filed is absent, we can obtain $w_d ^{\rm eff}<-1$, that
is we result to an effective dark energy presenting phantom
behaviour. This behaviour is pure effect of the interaction.
 \end{abstract}

\newpage

\section{Introduction}
Recent observations from type Ia supernovae \cite{SN} associated
with Large Scale Structure \cite{LSS} and Cosmic Microwave
Background anisotropies \cite{CMB} have provided main evidence for
the cosmic acceleration. The combined analysis of cosmological
observations suggests that the universe consists of about $70\%$
dark energy, $30\%$ dust matter (cold dark matter plus baryons), and
negligible radiation. Although the nature and origin of dark energy
are unknown, we still can propose some candidates to describe it,
namely  since we do not know where this dark energy comes from, and
how to compute it from the first principles, we search for
phenomenological models. The astronomical observations will then
select one of these models. The most obvious theoretical candidate
of dark energy is the cosmological constant $\lambda$ (or vacuum
energy) \cite{Einstein:1917,cc} which has the equation of state
parameter $w=-1$. However, as it is well known, there are two
difficulties that arise from the cosmological constant scenario,
namely the two famous cosmological constant problems --- the
``fine-tuning'' problem and the ``cosmic coincidence'' problem
\cite{coincidence}. An alternative proposal for dark energy is the
dynamical dark energy scenario. This  dynamical proposal is often
realized by some scalar field mechanism which suggests that the
specific energy form with negative pressure is provided by a scalar
field evolving down a proper potential. So far, a plethora of
scalar-field dark energy models have been studied, including
quintessence \cite{quintessence}, K-essence \cite{kessence}, tachyon
\cite{tachyon}, phantom \cite{phantom} and quintom \cite{quintom},
and so forth. An alternative way of explaining the observed
acceleration of the late universe is to modify the gravitational
theory and in the simplest case replace $R$ with $f(R)$ in the
action which is well known as $f(R)$ gravity. Here $f(R)$ is an
arbitrary function of scalar curvature (for recent reviews see
\cite{{12},{13}}). However, most of $f(R)$-gravity models do not
manage to pass the observational and theoretical tests (solar
system, neutron stars and binary pulsar constraints), giving also
rise to an unusual matter dominated epoch and leading to significant
fine-tunings \cite{koba}.
\\
Recently {Ho\v{r}ava proposed a renormalizable gravity theory with
higher spatial derivatives in four dimensions which reduces to
Einstein gravity with a non-vanishing cosmological constant in IR
but with improved UV behaviors \cite{{hor2},{hor1},{hor3}}. It is
similar to a scalar field theory of Lifshitz \cite{lif} in which the
time dimension has weight 3 if a space dimension has weight 1, thus
this theory is called {Ho\v{r}ava-Lifshitz gravity. This theory is
not invariant under the full diffeomorphism group of general
relativity, but rather under a subgroup of it, manifest in the
standard ADM splitting. The local symmetry usually puts constraint
on the system and only the physical modes appear as propagating
modes. However, since Ho\v{r}ava theory has not full diffeomorphism
invariance, one can not obtain full constraint to restrict the
possible modes to the physical modes. Various aspects of this theory
have been investigated
\cite{Calcagni:2009ar}-\cite{odint}.\\
Motivated by these, in the present work we are interested in
investigating interacting  dark energy model in the framework of
Ho\v{r}ava-Lifshitz cosmology.
\section{Ho\v{r}ava-Lifshitz cosmology with interacting  dark energy  }
In this section we obtain the equation of state for the dark energy
 when there is an interaction between  energy
density $\rho_d$ and a dark matter $\rho_m$. The dynamical variables
of Ho\v{r}ava-Lifshitz gravity are the lapse and shift functions,
$N$ and $N_i$ respectively, and the spatial metric $g_{ij}$.
Therefore we can write the metric as:
\begin{eqnarray}\label{1}
ds^2 = - N^2 dt^2 + g_{ij} (dx^i + N^i dt ) ( dx^j + N^j dt ) ,
\end{eqnarray}
The scaling transformation of the coordinates reads (z=3):
\begin{eqnarray}\label{2}
 t \rightarrow l^3 t~~~{\rm and}\ \ x^i \rightarrow l x^i~.
\end{eqnarray}
Decomposing the gravitational action into a kinetic and a potential
part as
\begin{equation}\label{3}
S_g = \int dt d^3x \sqrt{g} N ({\cal L}_K+{\cal L}_V)
\end{equation}
 and under the assumption of detailed
balance \cite{hor3}, which apart form reducing the possible terms in
the Lagrangian it allows for a quantum inheritance principle
\cite{hor2},
 the full action of Ho\v{r}ava-Lifshitz gravity is given by
\begin{eqnarray}\label{4}
&&S_g =  \int dt d^3x \sqrt{g} N \left\{ \frac{2}{\kappa^2}
(K_{ij}K^{ij} - \lambda K^2)
 - \frac{\kappa^2}{2 w^4} C_{ij}C^{ij}
 + \frac{\kappa^2 \mu}{2 w^2}
\frac{\epsilon^{ijk}}{\sqrt{g}} R_{il} \nabla_j R^l_k-
     \right. \nonumber \\
&&\left.   - \frac{\kappa^2 \mu^2}{8} R_{ij} R^{ij}  +
\frac{\kappa^2 \mu^2}{8(1 - 3 \lambda)} \left[ \frac{1 - 4
\lambda}{4} R^2 + \Lambda  R - 3 \Lambda ^2 \right] \right\}  \, ,
\end{eqnarray}
where
\begin{eqnarray}\label{5}
K_{ij} = \frac{1}{2N} \left( {\dot{g_{ij}}} - \nabla_i N_j -
\nabla_j N_i \right) \, ,
\end{eqnarray}
is the extrinsic curvature and
\begin{eqnarray}
\label{6}
 C^{ij} \, = \, \frac{\epsilon^{ijk}}{\sqrt{g}} \nabla_k
\bigl( R^j_i - \frac{1}{4} R \delta^j_i \bigr)
\end{eqnarray}
the Cotton tensor, and the covariant derivatives are defined with
respect to the spatial metric $g_{ij}$. $\epsilon^{ijk}$ is the
totally antisymmetric unit tensor, $\lambda$ is a dimensionless
constant and $\Lambda $ is a negative constant which is related to
the cosmological constant in the IR limit. Finally, the variables
$\kappa$, $w$ and $\mu$ are constants with mass dimensions $-1$, $0$
and $1$, respectively.\\
In order to add the matter component (including both dark and
baryonic matter)  in the theory one can follow two equivalent
approaches. The first is to introduce a scalar field
\cite{Calcagni:2009ar,Kiritsis:2009sh} and thus attribute to dark
matter a dynamical behavior, with its energy density $\rho_m$ and
pressure $p_m$ defined through the field kinetic and potential
energy. In the second approach one adds a cosmological
stress-energy tensor to the gravity field equations by demanding
to recover the usual general relativity formulation in the
low-energy limit \cite{Sotiriou:2009bx,Carloni:2009jc}. Similar
to \cite{Carloni:2009jc}, where the author add a matter sector to
the Ho\v{r}ava-Lifshitz action, we would like to add matter and
dark energy sectors with the following properties: It must
respect foliated diffeomorphism invariance, obey the principle of
detailed balance and be nontrivial at the $z=3$ critical point
and Lorentz invariant in the infrared. So inserting a scalar
field in the construction and imposing the corresponding
symmetries consistently, one results to the following action for
the matter field $\phi$ which has interaction with dark energy
field $\sigma$ (see \cite{Calcagni:2009ar,Kiritsis:2009sh} for
non-interacting case):
\begin{equation}\label{7}
S_m = \int dtd^3x \sqrt{g} N \left[
\frac{3\lambda-1}{4}\frac{\dot\phi^2}{N^2}
+m_1m_2\phi\nabla^2\phi-\frac{1}{2}m_2^2\phi\nabla^4\phi +
\frac{1}{2}m_3^2\phi\nabla^6\phi -V_t(\phi,\sigma)\right]~,
\end{equation}
where $V(\phi,\sigma)$ acts as a  total potential term \cite{maco}:
\begin{equation}\label{tot}
V_t(\phi,\sigma)=V(\phi)+B(\phi,\sigma),
\end{equation}
the interacting potential $B(\phi,\sigma)$ is a function ot both
fields, and $m_i$ are constants.

In usual Ho\v{r}ava-Lifshitz cosmological models, the scalar field
is responsible for dark matter. However, in principle one could
include additional scalars in the theory. In this work we will allow
for one more, in which we attribute the dark energy sector
\cite{sari}. Thus, we add a second scalar $\sigma$, with action
\begin{equation}\label{8}
S_d = \int dtd^3x \sqrt{g} N \left[
\frac{3\lambda-1}{4}\frac{\dot\sigma^2}{N^2}
+h_1h_2\sigma\nabla^2\sigma-\frac{1}{2}h_2^2\sigma\nabla^4\sigma +
\frac{1}{2}h_3^2\sigma\nabla^6\sigma -V_t(\phi,\sigma) \right]~,
\end{equation}
where $h_i$ are constants. Now, in order to focus on cosmological
frameworks, we impose an flat FRW metric,
\begin{eqnarray}\label{9}
N=1~,~~g_{ij}=a^2(t)\gamma_{ij}~,~~N^i=0~.
\end{eqnarray}
with
\begin{eqnarray}\label{10}
\gamma_{ij}dx^idx^j=dr^2+r^2d\Omega_2^2~.
\end{eqnarray}
We assume that the scalar fields are homogenous, i.e
$\phi\equiv\phi(t)$ and $\sigma\equiv\sigma(t)$. By varying $N$ and
$g_{ij}$, we obtain the equations of motion:
\begin{eqnarray}\label{Fr1}
H^2 =
\frac{\kappa^2}{6(3\lambda-1)}\left[\frac{3\lambda-1}{4}\,(\dot\phi^2+\dot\sigma^2)
+2V_t(\phi,\sigma)
 -\frac{3\kappa^2\mu^2\Lambda ^2}{8(3\lambda-1)}
 \right]
 \end{eqnarray}
\begin{eqnarray}\label{Fr2}
\dot{H}+\frac{3}{2}H^2 =
-\frac{\kappa^2}{4(3\lambda-1)}\left[\frac{3\lambda-1}{4}\,(\dot\phi^2+\dot\sigma^2)
-2V_t(\phi,\sigma) +\frac{3\kappa^2\mu^2\Lambda ^2}{8(3\lambda-1)}
 \right],
\end{eqnarray}
where we have defined the Hubble parameter as $H\equiv\frac{\dot
a}{a}$. Finally, the equations of motion for the scalar fields read:
\begin{eqnarray}\label{phidott}
&&\ddot\phi+3H\dot\phi+\frac{2}{3\lambda-1}(\frac{\partial V(\phi)}{\partial\phi}+
\frac{\partial B(\phi,\sigma)}{\partial\phi})=0\\
\label{sdott}
&&\ddot\sigma+3H\dot\sigma+\frac{2}{3\lambda-1}\frac{\partial
B(\phi,\sigma)}{\partial\sigma}=0.
\end{eqnarray}
Now we can define the energy density and pressure for the scalar
fields. Concerning the dark matter, the corresponding relations are
\begin{eqnarray}
&&\rho_m=\frac{3\lambda-1}{4}\,\dot\phi^2
+V(\phi)\nonumber\\
&&p_m=\frac{3\lambda-1}{4}\,\dot\phi^2 -V(\phi).
\end{eqnarray}
 Concerning the dark energy
sector, we have
\begin{equation}\label{rhoDE}
\rho_{d}\equiv\frac{3\lambda-1}{4}\,\dot\sigma^2 +B(\phi,\sigma)
-\frac{3\kappa^2\mu^2\Lambda ^2}{8(3\lambda-1)}
\end{equation}
\begin{equation}
\label{pDE} p_{d}\equiv\frac{3\lambda-1}{4}\,\dot\sigma^2
-B(\phi,\sigma) +\frac{3\kappa^2\mu^2\Lambda ^2}{8(3\lambda-1)}.
\end{equation}
The first parts of these expressions, namely
$\frac{3\lambda-1}{4}\,\dot\sigma^2B(\phi,\sigma) +$ and
$\frac{3\lambda-1}{4}\,\dot\sigma^2 -B(\phi,\sigma)$ correspond to
the energy density and pressure of the $\sigma$-field, $\rho_\sigma$
and $p_\sigma$ respectively. The last constant term is just the
explicit (negative) cosmological constant. Therefore, in expressions
(\ref{rhoDE}),(\ref{pDE}) we have defined the energy density and
pressure for the effective dark energy, which incorporates the
aforementioned contributions. Using the above definitions, we can
re-write the Friedmann equations (\ref{Fr1}),(\ref{Fr2}) in the
standard form:
\begin{eqnarray}
&&H^2 =
\frac{\kappa^2}{6(3\lambda-1)}\Big[\rho_M+\rho_{DE}\Big]\\
&&\dot{H}+\frac{3}{2}H^2 =
-\frac{\kappa^2}{4(3\lambda-1)}\Big[p_M+p_{DE}
 \Big].
\end{eqnarray}
Finally, note that using (\ref{phidott}),(\ref{sdott}) it is
straightforward to see that the aforementioned interacting dark
matter and dark energy quantities verify the following continuity
equations
\begin{eqnarray}\label{phidot2}
&&\dot{\rho}_m+3H(\rho_m+p_m)=\dot{\rho}_m+3H\rho_m(1+w_m)=-\dot{\phi}\frac{\partial
B(\phi,\sigma)}{\partial \phi}=Q\\ \label{sdot2}
&&\dot{\rho}_{d}+3H(\rho_{d}+p_{d})=dot{\rho}_{d}+3H\rho_{d}(1+w_{d})=\dot{\phi}\frac{\partial
B(\phi,\sigma)}{\partial \phi}=-Q.
\end{eqnarray}
The interaction is given by the quantity $Q=\Gamma \rho_{d}$. This
is a decaying of the dark energy component into dark matter with the
decay rate $\Gamma$. Taking a ratio of two energy densities as
$r=\rho_{\rm m}/\rho_{d}$, the above equations lead to
\begin{equation}
\label{2eq3} \dot{r}=3Hr\Big[w_{d}-+
\frac{1+r}{r}\frac{\Gamma}{3H}\Big]
\end{equation}
If we define \cite{Kim:2005at},
\begin{eqnarray}\label{eff}
w_d ^{\rm eff}=w_d+{{\Gamma}\over {3H}}\;, \qquad w_m ^{\rm
eff}=w_m-{1\over r}{{\Gamma}\over {3H}}\;.
\end{eqnarray}
Then, the continuity equations can be written in their standard
form
\begin{equation}
\dot{\rho}_\Lambda + 3H(1+w_d^{\rm eff})\rho_d =
0\;,\label{definew1}
\end{equation}
\begin{equation}
\dot{\rho}_m + 3H(1+w_m^{\rm eff})\rho_m = 0\; \label{definew2}
\end{equation}

Define as usual
\begin{equation} \label{2eq9} \Omega_{\rm
m}=\frac{\rho_{m}}{\rho_{cr}}=\frac{\kappa^2 \rho_{\rm
m}}{6(3\lambda-1)H^2},\hspace{1cm}\Omega_{\rm
d}=\frac{\rho_{d}}{\rho_{cr}}=\frac{\kappa^2 \rho_{\rm
d}}{6(3\lambda-1)H^2},
\end{equation}
Now we can rewrite the first Friedmann equation as
\begin{equation} \label{2eq10} \Omega_{\rm m}+\Omega_{\rm
d}=1.
\end{equation}
Using Eqs.(\ref{2eq9},\ref{2eq10}) we obtain following relation
for ratio of energy densities $r$ as
\begin{equation}\label{ratio}
r=\frac{1-\Omega_{d}}{\Omega_{d}}
\end{equation}

Here as in Ref.\cite{WGA}, we choose the following relation for
decay rate
\begin{equation}\label{decayeq}
\Gamma=3b^2(1+r)H
\end{equation}
with  the coupling constant $b^2$. Using Eq.(\ref{ratio}), the
above decay rate take following form
\begin{equation}\label{decayeq2}
\Gamma=\frac{3b^2H}{\Omega_{d}}
\end{equation}
Substitute this relation into Eq.(\ref{eff}), one find
\begin{equation} \label{3eq4}
w_d ^{\rm eff}=w_d+{b^2\over {\Omega_{d}}}\;, \qquad w_m ^{\rm
eff}=w_m-{b^2\over 1-\Omega_{d} }
\end{equation}
where
\begin{equation} \label{3eq41}
w_d=\frac{\frac{3\lambda-1}{4}\,\dot\sigma^2-B(\phi,\sigma)+\frac{3\kappa^2\mu^2\Lambda
^2}{8(3\lambda-1)}}{\frac{3\lambda-1}{4}\,\dot\sigma^2+B(\phi,\sigma)-\frac{3\kappa^2\mu^2\Lambda
^2}{8(3\lambda-1)}} ,\qquad
w_m=\frac{\frac{3\lambda-1}{4}\,\dot\phi^2-V(\phi)}{\frac{3\lambda-1}{4}\,\dot\phi^2+V(\phi)}
\end{equation}
Using Eqs.(\ref{rhoDE}), and (\ref{2eq9}) we obtain following
expression
\begin{equation} \label{3eq411}
w_d ^{\rm
eff}=\frac{\kappa^{2}(\frac{3\lambda-1}{4}\,\dot\sigma^2-B(\phi,\sigma)+\frac{3\kappa^2\mu^2\Lambda
^2}{8(3\lambda-1)})+6(3\lambda-1)b^2H^2}{\kappa^{2}(\frac{3\lambda-1}{4}\,\dot\sigma^2+B(\phi,\sigma)-\frac{3\kappa^2\mu^2\Lambda
^2}{8(3\lambda-1)})}
\end{equation}
If $\dot\sigma^2\leq \frac{-12b^2H^2}{\kappa^{2}}$, then $w_d
^{\rm eff}\leq -1$, therefore the interacting dark energy model
in the framework of Ho\v{r}ava gravity exhibiting phantom
behavior.\\
Now we consider the simplified case of the absence of the
$\sigma-$filed. In this case we have
\begin{equation} \label{3eq412}
w_d ^{\rm
eff}=\frac{-\kappa^{2}(B(\phi)-\frac{3\kappa^2\mu^2\Lambda
^2}{8(3\lambda-1)})+6(3\lambda-1)b^2H^2}{\kappa^{2}(B(\phi)-\frac{3\kappa^2\mu^2\Lambda
^2}{8(3\lambda-1)})}
\end{equation}
Considering the case $3\lambda-1>0$, if
$B(\phi)<\frac{3\kappa^2\mu^2\Lambda ^2}{8(3\lambda-1)}$, then
$w_d ^{\rm eff}<-1$. However in flat space non-interacting case,
where $b=0$, $w_d ^{\rm eff}=w_d =0$, so dark energy is a
cosmological constant and equation of state can not cross over
$-1$. This is big difference with the result of \cite{sari}.
Surprisingly in the absence of the $\sigma-$filed, when there is
an interaction between scalar matter filed with cosmological
constant we obtain $w_d ^{\rm eff}<-1$, that is we result to an
effective dark energy presenting phantom behaviour. This
behaviour is pure effect of the interaction, so we can ignore the
$\sigma-$filed at first and only allow an interaction between
matter filed $\phi$ and cosmological constant to obtain the
phantom-like behaviour. In this case the time evolution of
equation of state control by the dynamics of filed $\phi$.
 \section{Conclusions}
 In order to solve cosmological problems and because the lack of our
knowledge, for instance to determine what could be the best
candidate for dark energy to explain the accelerated expansion of
universe, the cosmologists try to approach to best results as
precise as they can by considering all the possibilities they have.
Studying the interaction between the dark energy and ordinary matter
will open a possibility of detecting the dark energy. It should be
pointed out that evidence was recently provided by the Abell Cluster
A586 in support of the interaction between dark energy and dark
matter \cite{Bertolami:2007zm}. However, despite the fact that
numerous works have been performed till now, there are no strong
observational bounds on the strength of this interaction
\cite{Feng:2007wn}. This weakness to set stringent (observational or
theoretical) constraints on the strength of the coupling between
dark energy and dark matter stems from our unawareness of the nature
and origin of dark components of the Universe. May be the recent
developments in Horawa gravity could offer a dark energy candidate
with perhaps better quantum gravitational foundations. So in the
present paper we have formulated Ho\v{r}ava-Lifshitz interacting
dark energy model. We have considered two scalar fields, one
responsible for dark matter fluid and one contributing to the dark
energy sector. Our calculations show that  the interacting dark
energy model in the framework of Ho\v{r}ava gravity exhibiting
phantom behavior.

\end{document}